# *Fermionic current in general relativity*

Elisa Varani

* Correspondence: elisa.varani@unicatt.it

**Abstract:**

This work aims to analyze the action of spinning particles in general relativity.

In general relativity the affine connection is required to be symmetric, so torsion is zero, we want to verify if a spinor field can be considered a torsion source. The broader reference context is Einstein Cartan's gravitational theory, which is a generalization of general relativity; in this theory in addition to curvature there is torsion associated with intrinsic angular momentum density. The affine connection is not restricted to be symmetric as required in general relativity, torsion is included due to the antisymmetric part of the affine connection. In Einstein Cartan's theory torsion is connected to the spin tensor as expressed by the Cartan equations. These equations are obtained through the variation of the total action with respect to the torsion; the total action is intended as the sum of the action for the gravitational field and the action for the fermionic field, $S = S_G + S_f$, according to the minimal action principle $\delta S = 0$.

We consider these important hints about torsion and spin tensor to revisit general relativity with spinor fields, we focus on the requirement of symmetric affine connection and develop the calculation of the spin coefficients.

In order to include fermions in general relativity we introduce a local reference frame and define a tetrad of basis vectors. We refer to the Hamiltonian formulation and calculate the canonical momenta associated with the temporal variation of the tetrads, we find a fermionic rotational term.

Starting from a torsion-less theory we get a rotational current that would generate a torsion contribution.

# 1 Introduction

## 1.1 Affine connection

This work moves between general relativity which is a torsion-free theory and the gravitational theory of Einstein- Cartan.

Torsion is a tensor defined from the affine connection $\Gamma^{\alpha}_{\beta\gamma}$:

$$T^{\alpha}_{\beta\gamma} = \Gamma^{\alpha}_{\beta\gamma} - \Gamma^{\alpha}_{\gamma\beta} \tag{1}$$

In general relativity the affine connection is symmetric, so torsion is zero.
The choice of the affine connection $\Gamma^{\alpha}_{\beta\gamma}$ is a consequence of two requirements:

- symmetry $\Gamma^{\alpha}_{\beta\gamma} = \Gamma^{\alpha}_{\gamma\beta}$
- metricity of the covariant derivative $\nabla_{\alpha} g_{\mu\nu} = 0$

Ref[3]

These conditions are fulfilled if the affine connection coincides with the metric connection defined by the Christoffel symbols

$$\Gamma^{\alpha}_{\beta\gamma} = \left\{ {\alpha \atop \beta\gamma} \right\} = \frac{1}{2} g^{\alpha\lambda}(\partial_{\beta} g_{\lambda\gamma} + \partial_{\gamma} g_{\lambda\beta} - \partial_{\lambda} g_{\beta\gamma}) \qquad (2)$$

**1. 2 Spinors in general relativity**

In general relativity, physical laws are required to maintain the same form under a general coordinate transformation (*Diff* $M^4$), according to the general covariance principle.
Mathematically this concept is expressed by Weyl's theorem which implies the choice of a symmetric metric connection for the description of the gravitational field.
Fermionic fields are described by spinors, spinors are a representation of the Lorentz group, there are no analogous objects for a general coordinate transformation.
Conforming to the equivalence principle it is possible to identify a system of inertial coordinates so that the effects of the gravitational field are canceled.
We consider a locally inertial reference frame, the tetrad field $e^{\mu}_a$ will be better defined later.

In order to construct the Dirac equation or more generally the action for the fermionic field we replace the ordinary derivative with a covariant derivative. The covariant derivative of a spinor must transform as a vector with respect to coordinate transformations and as a spinor with respect to a Lorentz transformation of the tetrad basis. Lorentz transformations rotate the vectors of the tetrad without changing the space-time coordinates
After these clarifications we write the generally covariant Dirac action and calculate canonical momenta in consonance with the Hamiltonian formulation.
This calculation leads to a fermionic rotational current.

**2 Materials and Method**

**2.1 Spinor action**

We have already mentioned tetrads as a local inertial frame (Cartan's repère mobile) [10].
The following formulas summarize the relations between tetrads and the metric:

$$g_{\mu\nu} = \eta_{ab} e^{a}_{\mu} e^{b}_{\nu} \qquad e^{a}_{\mu} e^{\nu}_{a} = \delta^{\nu}_{\mu} \qquad e^{\mu}_{a} e^{b}_{\mu} = \delta^{b}_{a} \qquad (3)$$

Both Greek and Latin indexes run from 0 to 3 and transform respectively under general coordinates and under local Lorentz transformations.

Within this system we consider fermionic fields.
The covariant derivative of a Dirac spinor $D_{\mu}\psi$ acts according to the following definitions:

$$D_\mu \psi = \partial_\mu \psi - \frac{i}{4} \omega_{\mu ab} \sigma^{ab} \psi \tag{4}$$

$$\overline{D_\mu \psi} = \partial_u \bar\psi + \frac{i}{4} \bar\psi \omega_{\mu ab} \sigma^{ab} \tag{5}$$

Where $\sigma^{ab} = \frac{i}{2}[\gamma^a, \gamma^b]$, $\gamma^a$ is usual (flat-space) $\gamma$-matrix and $\omega_{\mu ab}$ is the spin connection.

The spin coefficients are defined as follows:

$$\omega_{\mu ab} = e_{a\rho} \nabla_\mu e_b^\rho \tag{6}$$

$\nabla_\mu$ is the tensorial covariant derivative satisfying the metricity condition $\nabla_\mu g_{\nu\rho} = 0$

$$\omega_{\mu ab} = e_{a\rho} \nabla_\mu e_b^\rho = e_{a\rho} \left( \partial_\mu e_b^\rho + \Gamma^\rho_{\mu\lambda} e_b^\lambda \right) \tag{7}$$

The spin coefficients $\omega_{\mu ab}$ are antisymmetric in a and b,

$$\omega_{\mu ab} = -\omega_{\mu ba} \tag{8}$$

In the following calculations we use explicit antisymmetric equations (23)-(24).

We write the action for the fermionic field:

$$S(e_\mu^a, \psi, \bar\psi) = \frac{i}{2} \int d^4 x \sqrt{g} (\bar\psi \gamma^\mu D_\mu - \overline{D_\mu \psi} \gamma^\mu + 2im\bar\psi) \tag{9}$$

$\gamma^\mu = e_a^\mu \gamma^a$, and $d^4 x \sqrt{g}$ is the volume element

we make explicit the covariant derivative and write the action as follows:

$$S_f = \int d^4 x \sqrt{g} \times \left[ \frac{i}{2} \bar\psi \gamma^\mu \partial_\mu \psi - \frac{i}{2} \partial_u \bar\psi \gamma^\mu \psi + \frac{1}{8} \bar\psi \gamma^\mu \omega_{\mu ab} \sigma^{ab} \psi + \frac{1}{8} \bar\psi \sigma^{ab} \omega_{\mu ab} \gamma^\mu \psi - m \bar\psi \psi \right] \tag{10}$$

The terms of interaction with the spin connection are

$$\frac{1}{8} \bar\psi \gamma^\mu \omega_{\mu ab} \sigma^{ab} \psi + \frac{1}{8} \bar\psi \sigma^{ab} \omega_{\mu ab} \gamma^\mu \psi = \frac{1}{8} \bar\psi e_c^\mu \{\gamma^c, \sigma^{ab}\} \omega_{\mu ab} \psi \tag{11}$$

We define

$$T^{cab} = \{\gamma^c, \sigma^{ab}\} \tag{12}$$

The only nonvanishing elements of $T^{cab}$ are the following

$$T^{0ij} = \{\gamma^0, \sigma^{ij}\} = 2\varepsilon_{ijk} \begin{pmatrix} \sigma^k & 0 \\ 0 & -\sigma^k \end{pmatrix} \tag{13}$$

$$T^{i0j} = \{\gamma^i, \sigma^{0j}\} = -T^{0ij} \tag{14}$$

$$T^{nij} = \{\gamma^n, \sigma^{ij}\} = 2\varepsilon_{ijn} \begin{pmatrix} 0 & 1 \\ -1 & 0 \end{pmatrix} = 2\varepsilon_{ijn}\gamma^0\gamma^5 \tag{15}$$

The indices are flat indices: 0 timelike; n,i,j spacelike

**2 Materials and Methods**

**2.2 Canonical momenta**

We consider a global time function according to the 3+1 splitting of spacetime. Spacetime $(M^4, g_{\mu\nu})$ is assumed to be globally hyperbolic. Topologically this means $M^4 \approx R \times \Sigma$, $M^4$ admits regular foliation with non-intersecting three-dimensional space-like hypersurfaces $\Sigma_\tau$, $\tau$ is global time like function identifying the elements of the foliations (simultaneity surfaces).
We calculate the canonical moments according to the Hamiltonian formalism.

$$\pi_a^\alpha = \frac{\delta S_f}{\delta \partial_\tau e_a^\alpha}; \pi_0^\alpha = \frac{\delta S_f}{\delta \partial_\tau e_0^\alpha} \tag{16}$$

For further explanations, the reader could refer to [9].

**3. Results**

The only terms depending on the time derivative of the tetrad $\partial_\tau e_a^\alpha$ are related to the spin connection, we isolate and analyze only this part of the action (10).

$$\frac{1}{8}\bar{\psi}e_c^\mu T^{cab}\omega_{\mu ab}\psi \tag{17}$$

Since the only non-vanishing elements of $T^{cab}$ are $T^{0ij}, T^{nij}, T^{i0j}$ we only consider contributions arising from the terms

$$\frac{1}{8}\bar{\psi}e_0^\mu \omega_{\mu ij}T^{0ij}\psi \tag{18}$$

$$\frac{1}{8}\bar{\psi}e_i^\mu \omega_{\mu 0 j} T^{i0j}\psi \qquad (19)$$

$$\frac{1}{8}\bar{\psi}e_n^\mu \omega_{\mu ij} T^{nij}\psi \qquad (20)$$

Spin coefficients are depending on tetrads, we make explicit calculations

$$\omega_{\mu ij} = e_{i\rho}\nabla_\mu e_j^\rho = e_{i\rho}\left(\partial_\mu e_j^\rho + \Gamma^\rho_{\mu\lambda}e_j^\lambda\right) \qquad (21)$$

$$\omega_{\mu 0 j} = e_{0\rho}\nabla_\mu e_j^\rho = e_{0\rho}\left(\partial_\mu e_j^\rho + \Gamma^\rho_{\mu\lambda}e_j^\lambda\right) \qquad (22)$$

The same we do for Christoffel symbols,

$$\Gamma^\alpha_{\beta\gamma} = \left\{{\alpha \atop \beta\gamma}\right\} = \frac{1}{2}g^{\alpha\lambda}(\partial_\beta g_{\lambda\gamma} + \partial_\gamma g_{\lambda\beta} - \partial_\lambda g_{\beta\gamma})$$

according to the equation $g_{uv} = e_\mu^a e_{va}$ we replace the metric tensor with tetrads and look for $\partial_\tau e_a^\alpha$.

Spin coefficients are antisymmetric, so we write

$$\omega_{\mu ij} = \frac{1}{2}\left(e_{i\rho}\partial_\mu e_j^\rho - e_{j\rho}\partial_\mu e_i^\rho\right) + \frac{1}{2}\Gamma^\rho_{\mu\lambda}(e_{i\rho}e_j^\lambda - e_{j\rho}e_i^\lambda) \qquad (23)$$

$$\omega_{\mu 0 j} = \frac{1}{2}\left(e_{0\rho}\partial_\mu e_j^\rho - e_{j\rho}\partial_\mu e_0^\rho\right) + \frac{1}{2}\Gamma^\rho_{\mu\lambda}(e_{0\rho}e_j^\lambda - e_{j\rho}e_0^\lambda) \qquad (24)$$

Temporal derivatives of the tetrad are manifest in these parts:

$$\omega_{\tau ij} = \frac{1}{2}\left(e_{i\rho}\partial_\tau e_j^\rho - e_{j\rho}\partial_\tau e_i^\rho\right) \qquad (25)$$

$$\omega_{\tau 0 j} = \frac{1}{2}\left(e_{0\rho}\partial_\tau e_j^\rho - e_{j\rho}\partial_\tau e_0^\rho\right) \qquad (26)$$

The terms including time derivatives of the tetrad present in the Christoffel symbols are cancelled out, due to the antisymmetric part of the spin coefficients.

We only consider these contributions

$$\frac{1}{8}\bar{\psi}e_0^\tau \omega_{\tau ij}T^{0ij}\psi = \frac{1}{8}\bar{\psi}e_0^\tau[\frac{1}{2}\left(e_{i\rho}\partial_\tau e_j^\rho - e_{j\rho}\partial_\tau e_i^\rho\right)]T^{0ij}\psi \tag{27}$$

$$\frac{1}{8}\bar{\psi}e_i^\tau \omega_{\tau 0j}T^{i0j}\psi = \frac{1}{8}\bar{\psi}e_i^\tau[\frac{1}{2}\left(e_{0\rho}\partial_\tau e_j^\rho - e_{j\rho}\partial_\tau e_0^\rho\right)]T^{i0j}\psi \tag{28}$$

$$\frac{1}{8}\bar{\psi}e_n^\tau \omega_{\tau ij}T^{nij}\psi = \frac{1}{8}\bar{\psi}e_n^\tau\left[\frac{1}{2}\left(e_{i\rho}\partial_\tau e_j^\rho - e_{j\rho}\partial_\tau e_i^\rho\right)\right]T^{nij}\psi \tag{29}$$

Finally we obtain the conjugated momenta:

$$\pi_a^\alpha = \frac{\delta S_f}{\delta \partial_\tau e_a^\alpha} = \frac{1}{8}\bar{\psi}\left[e_0^\tau e_{i\alpha}T^{0ia} + \frac{1}{2}e_i^\tau e_{0\alpha}T^{i0a} + e_n^\tau e_{i\alpha}T^{nia}\right]\psi \tag{30}$$

Index a is flat spacelike

$$\pi_0^\alpha = \frac{\delta S_f}{\delta \partial_\tau e_0^\alpha} = \frac{1}{8}\bar{\psi}\left[-\frac{1}{2}e_i^\tau e_{j\alpha}T^{i0j}\right]\psi \tag{31}$$

Index 0 is flat timelike

The term (28) provides two contributions.

**3.1 Tetrads in the Schwinger gauge**

Consider the Schwinger time gauge condition $e_r^0 = 0$, we can express tetrads as follows:

$e_\tau^0 = N; e_r^0 = 0$

$e_\tau^a = N^r e_r^a = N^a; e_r^a = e_r^a$ \hfill (32)

Coterads are so defined

$e_0^\tau = \frac{1}{N}; e_n^\tau = 0$

$e_0^r = -\frac{N^r}{N}; e_n^r = e_n^r$ \hfill (33)

Canonical momenta become:

$$\pi_a^r = \frac{\delta S_f}{\delta \partial_\tau e_a^r} = \frac{1}{8}\bar{\psi}[\frac{1}{N}e_{ir}T^{0ia}]\psi; \quad \alpha = r \tag{34}$$

$$\pi_0^\alpha = \frac{\delta S_f}{\delta \partial_\tau e_0^\alpha} = 0 \tag{35}$$

**4. Summury and conclusion**

The last considerations concern the analysis of interactions and the comparison with the results obtained in linearized general relativity [7]

We write the Dirac Lagrangian

$$L_f = \frac{i}{2}\bar{\psi}\gamma^\mu \partial_\mu \psi - \frac{i}{2}\partial_u \bar{\psi}\gamma^\mu \psi + \frac{1}{8}\bar{\psi}e_c^\mu \{\gamma^c, \sigma^{ab}\}\omega_{\mu ab}\psi - m\bar{\psi}\psi \tag{36}$$

since $\gamma^\mu = e_a^\mu \gamma^a$, (36) becomes

$$L_f = \frac{i}{2}e_a^\mu \overline{(\psi}\gamma^a \partial_\mu \psi - \partial_u \bar{\psi}\gamma^a \psi) + \frac{1}{8}\bar{\psi}e_c^\mu \{\gamma^c, \sigma^{ab}\}\omega_{\mu ab}\psi - m\bar{\psi}\psi \tag{37}$$

We express the Dirac Lagrangian replacing equations of $T_\mu^a$ and $s^{cab}$,

$T_\mu^a = \overline{(\psi}\gamma^a \partial_\mu \psi - \partial_u \bar{\psi}\gamma^a \psi)$ energy momentum tensor

$s^{cab} = -\frac{i}{4}\bar{\psi}\{\gamma^c, \sigma^{ab}\}\psi$ spin angular momentum tensor

$$L_f = \frac{i}{2}e_a^\mu T_\mu^a + \frac{i}{2}e_c^\mu \omega_{\mu ab} s^{cab} - m\bar{\psi}\psi \tag{38}$$

There are two terms of interaction between the gravitational field and the fermionic field.

The first interaction term $e_a^\mu T_\mu^a$ consists of the product of the tetrad $e_a^\mu$ with the energy-momentum tensor $T_\mu^a$ for the Dirac field, the appearance of the tetrad field $e_a^\mu$ is due to modified $\gamma$-matrix being $\gamma^\mu = e_a^\mu \gamma^a$. This interaction is also evident in linearized gravity.

The second term $e_c^\mu \omega_{\mu ab} s^{cab}$ is related to the spin connection; in weak gravity the interaction part containing the spin connection vanishes (only to first order).

The linearized metric tensor is obtained by the flat metric plus a small perturbation $h_{\mu\nu}$

$$g_{\mu\nu} = \eta_{\mu\nu} + h_{\mu\nu} \tag{39}$$

In the linearized theory of general relativity, the interaction is represented by the product of the field $h_{\mu\nu}$ with the Dirac energy- momentum tensor $T^{\mu\nu}$.

In the non-relativistic limit this interaction has gravitoelectric and gravitomagnetic effects, both orbital and spin angular momenta couple in the same way to the gravitomagnetic field, the precession rate is universal for any angular systems [7].

Gravitomagnetism seems well explained with linearized gravity.

In this work we have analyzed in detail the spin coefficients.

Momenta $\pi_a^\alpha$ and $\pi_0^\alpha$ have been calculated from (11), this term disappears if we consider the linearized theory for gravity, as already said.

Canonical momenta are related to the spin rotational current; we can compare the second member of the equations (30); (31) with the canonical spin angular momentum tensor.

$$s^{abc} = -\frac{i}{4}\bar{\psi}\{\gamma^a, \sigma^{bc}\}\psi \tag{40}$$

The results we obtained confirm the interaction between the fermionic field and the gravitational field, moreover the spinor rotational current, usually associated with torsion, is highlighted through the calculation of canonical momenta.

**Appendix A: Tetrads and cotetrads**

Let $(M^4, g_{\mu\nu})$ be a spacetime, consider a generic point $p \in M^4$, $x^\mu(p)$ are local coordinates.
We consider a transformation that leads to inertial local coordinates $x^\mu \to X^a$.
The flat metric in $p$ is expressed by the law of transformation of tensors under coordinate change

$$\eta_{ab}(p) = \frac{\partial X^a}{\partial x^\mu}\frac{\partial X^b}{\partial x^\nu} g_{\mu\nu} \tag{A.1}$$

We define tetrads as follows

$$e_a^\mu(p) = \frac{\partial x^\mu(p)}{\partial X^a} \tag{A.2}$$

$$\eta_{ab}(p) = e_a^\mu e_b^\nu g_{\mu\nu} \tag{A.3}$$

The 4-metric can be expressed in terms of orthonormal cotetrads or local coframes

$$e_\mu^a(p) = \frac{\partial X^a(p)}{\partial x^\mu} \tag{A.4}$$

$$g_{\mu\nu}(p) = \frac{\partial X^a}{\partial x^\mu}\frac{\partial X^b}{\partial x^\nu}\eta_{ab} \qquad (A.5)$$

$$g_{\mu\nu}(p) = e_\mu^a e_\nu^b \eta_{ab} \qquad (A.6)$$

$$e_\mu^a e^{\nu b} = \eta^{ab} \qquad e_a^\mu e_{\mu b} = \eta_{ab}$$

## Appendix A: *lapse e shift* function

Let $M^4$ be a globally hyperbolic manifold, consider splitting 3+1 with foliation surfaces $\Sigma_\tau$, $x^\mu = z^\mu(\tau,\sigma)$ represents the coordinates of the points of $\Sigma_\tau$; $\{\vec{\sigma}\}$ are local 3 coordinates on $\Sigma_\tau$; $l^\mu(\tau,\sigma)$ is the unit controvariant vector normal to $\Sigma_\tau$ at $\vec{\sigma}$ . Ref[9].

we can thus define the lapse and shift functions:

-The positive function $N(\sigma) > 0$ is the lapse function:

$N(\sigma)d\tau$ measures the proper time interval between the two hypersurfaces $\Sigma_\tau$ e $\Sigma_{\tau+d\tau}$

-$N^r(\sigma)$ is the shift function , $N^r(\sigma)\, d\tau$ describes the horizontal shift

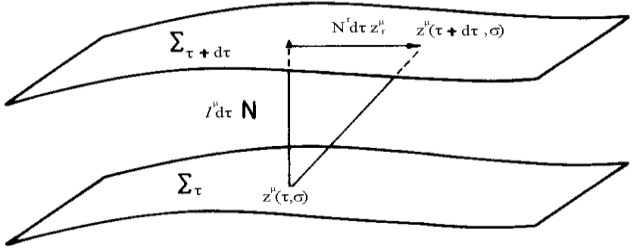

Figure 1: lapse and shift

**Acknowledgments**

I wish to thank Ronald Adler.

**Data availability**

I have no data associated with the manuscript, as the article is about theoretical analysis.

**References and further readings**